\documentclass{PoS}
\rightline{\parbox{3cm}{\large\rm
DESY 11-234
}}

\usepackage{amsmath}
\usepackage{subfig}

\newcommand{\dd}{{\rm d}}

\title{An Update on Distribution Amplitudes of the Nucleon and its Parity Partner}

\ShortTitle{An Update on Distribution Amplitudes of the Nucleon and its Parity Partner}

\author{\speaker{R.~W.~Schiel} $^{,a}$, G.~S.~Bali $^a$, V.~M.~Braun $^a$, S.~Collins $^a$, M.~G\"ockeler $^a$, C.~Hagen $^a$, R.~Horsley $^b$, Y.~Nakamura $^c$, D.~Pleiter $^{d,a}$, P.~E.~L.~Rakow $^e$, A.~Sch\"afer $^a$, G.~Schierholz $^f$, H.~St\"uben $^g$, P.~Wein $^a$, J.~M.~Zanotti $^b$ \\
        \llap{$^a$} Institut f\"ur Theoretische Physik, Universit\"at Regensburg, 93040 Regensburg, Germany\\
        \llap{$^b$} School of Physics and Astronomy, University of Edinburgh, Edinburgh EH9 3JZ, UK \\
        \llap{$^c$} RIKEN Advanced Institute for Computational Science, Kobe, Hyogo 650-0047, Japan \\
        \llap{$^d$} JSC, J\"ulich Research Centre, 52425 J\"ulich, Germany \\
        \llap{$^e$} Theoretical Physics Division, Department of Mathematical Sciences, University of Liverpool, Liverpool L69 3BX, UK \\
        \llap{$^f$} Deutsches Elektronen-Synchrotron DESY, 22603 Hamburg, Germany \\
        \llap{$^g$} Konrad-Zuse-Zentrum f\"ur Informationstechnik Berlin, 14195 Berlin, Germany \\
        E-mail: \email{rainer.schiel@physik.uni-regensburg.de}
	\begin{center} (QCDSF Collaboration) \end{center}}

\abstract{The calculation of baryon wave functions at small inter-quark separations is an ongoing effort within the QCDSF collaboration \cite{arXiv:0811.2712, arXiv:0902.3087, arXiv:1011.1092}. In this update on normalization constants and distribution amplitudes of the nucleon and its negative parity partner, $N^* (1535)$, we present new lattice data which helps us controlling finite size effects. We use new chiral perturbation theory results to perform the extrapolation to the physical point.}

\FullConference{XXIX International Symposium on Lattice Field Theory \\
		 July 10-16 2011\\
		 Squaw Valley, Lake Tahoe, California}

\begin{document}

\section{Introduction}

Distribution amplitudes are wave functions of quarks and gluons within a hadron at small transverse separations between the constituents. They are particularly useful for the calculation of hard exclusive reactions at large momentum transfer. In this case, it is sufficient to consider the valence Fock state, i.e.\ the three-quark state for baryons. We have computed the normalization constants and moments of the distribution amplitude of both the nucleon and its negative parity partner, the $N^* (1535)$, from first principles in lattice QCD. An update on these calculations is presented below.

In the continuum, in the infinite-momentum-frame, the leading-twist component of the nucleon (and similarly the $N^* (1535)$) wave function can be written as
\begin{equation}
| N, \uparrow \rangle = f_N \int \frac{ [\dd x] \varphi (x_i)}{2 \sqrt{24 x_1 x_2 x_3}} \{ | u^{\uparrow} (x_1) u^{\downarrow} (x_2) d^{\uparrow} (x_3) \rangle -  | u^{\uparrow} (x_1) d^{\downarrow} (x_2) u^{\uparrow} (x_3) \rangle \}.
\label{eqwavefun} \end{equation}
Here, the arrows indicate the spin, the transverse momentum components have been integrated out, $x_i$ are the longitudinal momentum fractions, integration is defined by $\int[dx] \equiv \int_0^1 \dd x_1 \dd x_2 \dd x_3 \delta(1-x_1-x_2-x_3) $, $f_N$ is the normalization constant which corresponds to the ``wave function at the origin'', and $\varphi (x_i)$ is the leading-twist nucleon distribution amplitude (NDA): it contains information about the distribution of the momenta of the three valence quarks.

The NDA can be expanded into (one-loop) multiplicatively renormalizable terms \cite{arXiv:0806.2531},
\[
\begin{split}
\varphi (x_i; \mu^2) = & 120 x_1 x_2 x_3 \Big\{1 + c_{10} (x_1 - 2 x_2 +x_3) L^{\frac{8}{3\beta_0}} + c_{11} (x_1 - x_3) L^{\frac{20}{9\beta_0}} \\
& + c_{20} \left[ 1 + 7 (x_2 - 2 x_1 x_3 - 2 x_2^2) \right] L^{\frac{14}{3\beta_0}} + c_{21} \left( 1 - 4 x_2 \right) \left( x_1 - x_3 \right) L^{\frac{40}{9\beta_0}} \\ 
& \left. + c_{22} \left[ 3 - 9 x_2 + 8 x_2^2 - 12 x_1 x_3 \right] L^{\frac{32}{9 \beta_0}}+ \ldots \right\},
\end{split}
\]
where $L \equiv \alpha_s (\mu) / \alpha_s (\mu_0)$ and the $c_{ij}$ are the so-called ``shape parameters'' at a scale $\mu_0$. In the asymptotic case $\mu \rightarrow \infty$, all shape parameters are zero. The shape parameters of first and second order, $c_{1j}$ and $c_{2j}$, are directly related to the first and second moments of the distribution amplitude, $\varphi^{lmn}$ (with $l+m+n = 1$ or $2$). The moments of the NDA are defined by
\[
\varphi^{lmn} = \int [\dd x] x_1^l x_2^m x_3^n \varphi (x_1, x_2, x_3).
\]
Momentum conservation ($x_1 + x_2 + x_3 = 1$) demands that the $\varphi^{lmn}$ fulfill the constraints
\begin{equation}
\varphi^{lmn} = \varphi^{(l+1)mn} + \varphi^{l(m+1)n} + \varphi^{lm(n+1)},
\label{momconst} \end{equation}
where $\varphi^{000} \equiv 1$. The five shape parameters of first and second order and four constraints of the form eq.~(\ref{momconst}) uniquely determine the nine first and second order moments, and vice versa.

The next-to-leading twist components of the nucleon wave function provide information about the orbital angular momentum of the quark in the nucleon. In this work, we have only determined the corresponding normalization constants $\lambda_1$ and $\lambda_2$.  

\section{Distribution Amplitudes from Lattice QCD}

To determine the NDAs from lattice QCD, several tasks have to be performed. Irreducibly transforming multiplets of three-quark operators have been worked out in \cite{arXiv:0801.3932} and are shown in Table \ref{irred}. The renormalization constants were computed in \cite{arXiv:0810.3762}. These are based on a non-perturbative renormalization and subsequent 1-loop-conversion to the $\overline{\rm MS}$ scheme. The calculation of 2-loop conversion factors, based on a consistent subtraction scheme for three-quark operators in dimensional regularisation, is currently in progress.

\begin{table}
\centering
\begin{tabular}{|c||c|c|c|}
\hline & dimension 9/2 & dimension 11/2 & dimension 13/2 \\
  & (0 derivatives)   & (1 derivatives)   & (2 derivatives) \\
\hline \hline $\tau^{\underline{4}}_1$ & $\mathcal{B}^{(0)}_{1,i}, \mathcal{B}^{(0)}_{2,i}, \mathcal{B}^{(0)}_{3,i}, \mathcal{B}^{(0)}_{4,i}, \mathcal{B}^{(0)}_{5,i}$ & & $\mathcal{B}^{(2)}_{1,i},\mathcal{B}^{(2)}_{2,i},\mathcal{B}^{(2)}_{3,i}$ \\
\hline $\tau^{\underline{4}}_2$ & & & $\mathcal{B}^{(2)}_{4,i},\mathcal{B}^{(2)}_{5,i},\mathcal{B}^{(2)}_{6,i}$ \\
\hline $\tau^{\underline{8}}$ & $\mathcal{B}^{(0)}_{6,i}$ & $\mathcal{B}^{(1)}_{1,i}$ & $\mathcal{B}^{(2)}_{7,i},\mathcal{B}^{(2)}_{8,i},\mathcal{B}^{(2)}_{9,i}$ \\
\hline $\tau^{\underline{12}}_1$ & $\mathcal{B}^{(0)}_{7,i}, \mathcal{B}^{(0)}_{8,i}, \mathcal{B}^{(0)}_{9,i}$ & $\mathcal{B}^{(1)}_{2,i}, \mathcal{B}^{(1)}_{3,i}, \mathcal{B}^{(1)}_{4,i}$ & $\mathcal{B}^{(2)}_{10,i},\mathcal{B}^{(2)}_{11,i},\mathcal{B}^{(2)}_{12,i},\mathcal{B}^{(2)}_{13,i}$ \\
\hline $\tau^{\underline{12}}_2$ & & $\mathcal{B}^{(1)}_{5,i}, \mathcal{B}^{(1)}_{6,i}, \mathcal{B}^{(1)}_{7,i}, \mathcal{B}^{(1)}_{8,i}$ & $\mathcal{B}^{(2)}_{14,i},\mathcal{B}^{(2)}_{15,i},\mathcal{B}^{(2)}_{16,i},\mathcal{B}^{(2)}_{17,i},\mathcal{B}^{(2)}_{18,i}$ \\ \hline
\end{tabular}
\caption{Multiplets of three-quark operators, grouped by dimension and irreducible representation \cite{arXiv:0801.3932}.} \label{irred}
\end{table}

On the lattice, correlation functions of the form
\[
\langle \mathcal {O}(x)_{\alpha \beta \gamma} \bar{\mathcal{N}} (y)_\tau \rangle
\]
have been calculated. Here, $\mathcal{N}$ is a smeared nucleon interpolator and $\mathcal{O}$ is a local three-quark operator with up to two derivatives. $\mathcal{O}$ belongs to one of the irreducibly transforming multiplets of three-quark operators. The separation of positive and negative parity states, i.e. of the nucleon and the $N^*(1535)$, has been achieved with the generalized Lee-Leinweber parity ``projector'' $\frac{1}{2} (1 \pm \frac{m}{E} \gamma_4)$ \cite{hep-lat/9809095}.

The normalization constants $f_N$ (or $f_{N^*}$), $\lambda_1$ and $\lambda_2$ are obtained from $\mathcal{O}$s without derivatives. Out of the several possible operators (see Table \ref{irred}), those which are convenient to calculate have been chosen. The first and second moments of the distribution amplitude require $\mathcal{O}$s with one and two derivatives, respectively. In order to avoid mixing with operators of lower dimension, one is limited to one-derivative operators in $\tau^{\underline{12}}_2$ and two-derivative operators in $\tau^{\underline{4}}_2$. Operators with three or more derivatives, which would yield higher moments of the NDA, are excluded from this study. The first and second order shape parameters are then determined from a fit of the first and second moments, constrained by eq.~(\ref{momconst}).

The lattices that we have used to determine the NDAs are shown in Table \ref{latti}. The three volumes at $\beta=5.29, \kappa=0.13632$ provide good insight into finite volume effects. Also, a new lattice at a pion mass of less than $200\ \rm{MeV}$ has been analyzed. This brings us closer to the physical point, making extrapolations easier. Unfortunately, we have not analyzed lattices with reasonably low pion masses at lattice spacings other than $a \approx 0.072\ \rm{fm}$ ($\beta = 5.29$) yet. So, while the larger mass data of the $\beta = 5.40$ lattices do not indicate significant discretization effects, there is still a remaining continuum extrapolation uncertainty at small quark masses.

\begin{table}
\centering
\begin{tabular}{|c|c|c|c|}
\hline
$\kappa$ & $m_\pi /$ MeV & Size & \# Configurations\\
\hline \hline
\multicolumn{4}{|c|}{$\beta = 5.29, a = 0.0716$ fm} \\
\hline 0.13590 & 660$^\dagger$ & $24^3 \times 48$ & 901 \\
\hline 0.13620 & 428 & $24^3 \times 48$ & 850 \\
\hline 0.13632 & 306 & $24^3 \times 48$ & 540 \\
\hline 0.13632 & 295 & $32^3 \times 64$ & 950 \\
\hline 0.13632 & 288 & $40^3 \times 64$ & 858 \\
\hline 0.13640 & 182 & $48^3 \times 64$ & 798$^*$ \\
\hline \hline \multicolumn{4}{|c|}{$\beta = 5.40, a = 0.0604$ fm} \\
\hline 0.13610 & 722$^\dagger$ & $24^3 \times 48$ & 687 \\
\hline 0.13625 & 622$^\dagger$ & $24^3 \times 48$ & 1180 \\
\hline 0.13640 & 503 & $24^3 \times 48$ & 1037 \\
\hline
\end{tabular}
\caption{Lattice parameters. The pion mass is at the given lattice volume, i.e.\ not extrapolated to infinite volume. $^*$~Results from this lattice are to be considered preliminary. $^\dagger$~The pion mass of these lattices is beyond the range of applicability of chiral perturbation theory; therefore, these lattices were not used to fit the $\chi$PT low-energy constants.} \label{latti}
\end{table}

For the normalization constants and distribution amplitudes of the nucleon, we have results from $1$-loop covariant baryon chiral perturbation theory to improve both the extrapolation to the physical pion mass and to infinite volume \cite{arXiv:1106.3440}. For each observable, there are two low energy constants that we fit with lattice data. For the $N^* (1535)$, however, results from chiral perturbation theory do not exist as yet.

\section{Results}

\begin{figure}
\begin{center}
\subfloat[]{
\includegraphics[width=0.44\textwidth]{fNnewProc.eps}
\label{fNNstarplot}\hspace{0.3cm}}
\subfloat[]{
\includegraphics[width=0.52\textwidth]{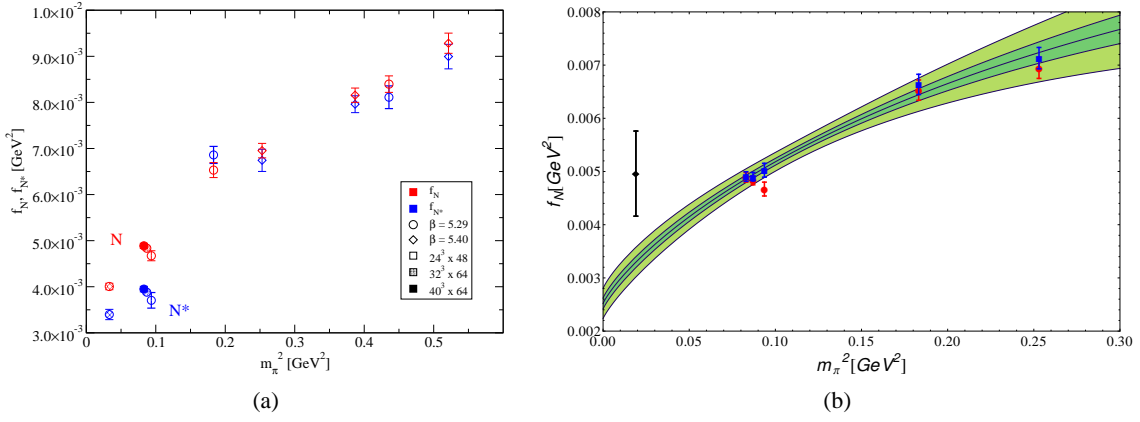}
\label{fNchiralplot}}
\end{center}
\caption{The leading-twist normalization constants $f_N$ and $f_{N^*}$ as functions of the pion mass. The chiral extrapolation of $f_N$ is shown in (b). The green bands are the $1$- and $2$-$\sigma$ errors of the chiral extrapolation. The red points are the lattice values and the blue points are ``finite volume corrected'' values. The left most (black) data point shows the value for $f_N$ from QCD sum rule calculations \cite{arXiv:1011.0758}.} \label{fNplot}
\end{figure}

In the following, we show selected results. The leading-twist normalization constants $f_N$ and $f_{N^*}$ are displayed in Figure \ref{fNNstarplot}. While $f_N$ and $f_{N^*}$ are -- within the error bars -- almost identical at higher pion masses, they are significantly different from each other at $m_\pi \lesssim 300\ \rm{MeV}$: the nucleon's leading-twist wave function at the origin is bigger than that of its parity partner. 

The chiral perturbation theory fit to $f_N$ is shown in Figure \ref{fNchiralplot}. Once the finite volume corrections have been applied, the lattice data points lie nicely on the fitted curve.

\begin{figure}
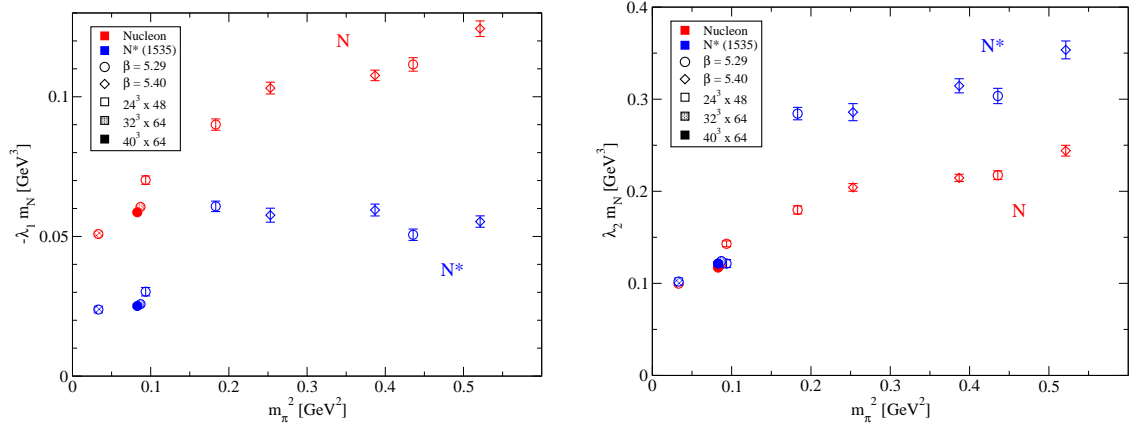

\begin{center}
\subfloat{
\includegraphics[width=0.47\textwidth]{lambda1newProc.eps}
\hspace{.50cm}}
\subfloat{
\includegraphics[width=0.47\textwidth]{lambda2newProc.eps}}
\end{center}
\caption{The next-to-leading twist normalization constants $\lambda_1$ (left) and $\lambda_2$ (right) as a function of the pion mass.} \label{lambdasplot}
\end{figure}

In Figure~\ref{lambdasplot}, the next-to-leading twist normalization constants, $\lambda_1$ and $\lambda_2$, of the nucleon and the $N^* (1535)$ can be seen. For the nucleon, we find that $\lambda_2 \approx - 2 \lambda_1$, which should hold exactly in the non-relativistic limit. The $N^* (1535)$ shows a curious behavior towards lower pion masses, for which we do not have a conclusive explanation as yet. 

\begin{figure}
\centering
\includegraphics[width=0.6\textwidth]{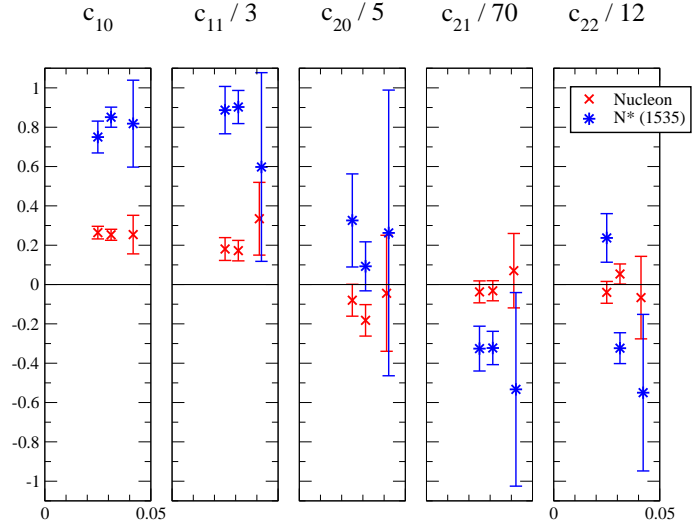}
\caption{Volume dependence of the shape parameters at $\beta = 5.29$ and $\kappa = 0.13632$. Shown on the $x$-axis is the inverse length of the lattice $1/L$, where $L=24, 32, 40$.} \label{shapevsvol}
\end{figure}

In Figure \ref{shapevsvol}, we show the volume dependence of the shape parameters. No statistically significant effects can be seen in most cases. One exception is $c_{22}$ of the $N^* (1535)$, where a larger deviation between the $32^3 \times 64$ and $40^3 \times 64$ lattices occurs but, given the relatively stable behavior of the other shape parameters, we attribute this to a larger-than-normal statistical fluctuation.  

The first-order shape parameters $c_{1i}$ of the $N^* (1535)$ are much bigger than those of the nucleon, indicating a stronger deviation from the asymptotic wave function. The $c_{2i}$, however, (maybe with the exception of the $N^*(1535)$ $c_{21}$) are consistent with zero, within the error bars. 

\begin{figure}
\centering
\includegraphics[width=0.9\textwidth]{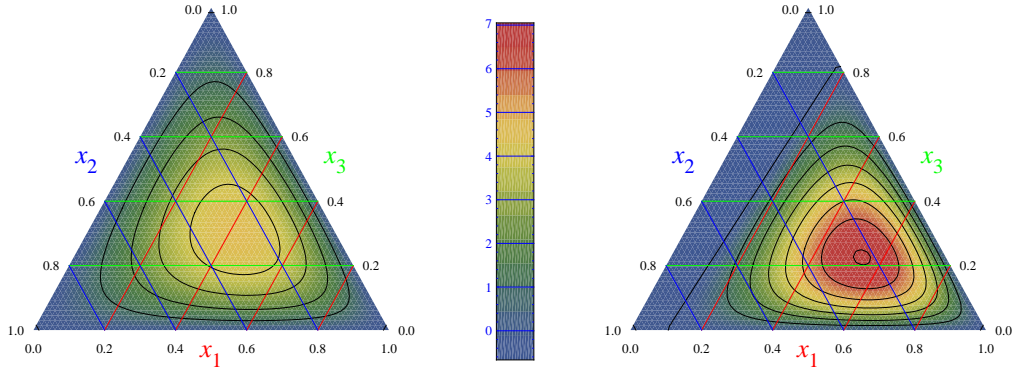}
\caption{Barycentric plot of the nucleon (left) and the $N^* (1535)$ (right) wave functions. Due to the large error of the second moments, only the first moments are shown in this plot.} \label{barycent}
\end{figure}

Using the first order shape parameters of the $\beta = 5.29, \kappa = 0.13632, 40^3 \times 64$ lattice, we have created the barycentric plots of the wave functions in Figure \ref{barycent}. At asymptotically large momentum transfers, the wave functions will be completely symmetric. A shift of the maximum of the nucleon wave function towards higher values of $x_1$ (see Eq.~(\ref{eqwavefun})) can be clearly seen and, in the case of the $N^* (1535)$, this shift is even more pronounced. 

\section{Conclusions}

The wave function normalization constants and low moments of the leading twist distribution amplitudes of the nucleon and $N^* (1535)$ have been calculated from lattice QCD. The set of lattices that were used spans a wide range of pion masses and lattice volumes. Extrapolations to the physical pion mass and infinite volume were made, using chiral perturbation theory formulae, in the case of the nucleon. 

\begin{acknowledgments}

This work has been supported in part by the Deutsche Forschungsgemeinschaft (SFB/TR 55) and the European Union under Grant Agreement number 238353 (ITN STRONGnet). SC acknowledges support from the Claussen-Simon-Foundation (Stifterverband f\"ur die Deutsche Wissenschaft). The computations were performed on the QPACE systems of the SFB/TR 55, Regensburg's Athene HPC cluster and J\"ulich's JUGENE using the Chroma software system \cite{hep-lat/0409003}.

\end{acknowledgments}

\end{document}